\documentclass[pdflatex,twocolumn,sn-nature,super]{sn-jnl}

\usepackage{silence}
\usepackage{preamble}

\setabbreviationstyle[species]{long-em-short-em}
\glssetcategoryattribute{species}{nohyper}{true}
\newabbreviation[category={species}]{celegans}{C. elegans}{Caenorhabditis elegans}
\newabbreviation[category={species}]{dmelanogaster}{D. melanogaster}{Drosophila melanogaster}

\newcommand{\im}{\mathrm{i}}

\DeclareCaptionLabelSeparator{bar}{|}
\defasforeign[aka]{a.k.a.} 

\pgfplotsset{simulation scatter/.style={only marks, blue}}
\pgfplotsset{full theory line/.style={orange}}
\pgfplotsset{approx theory line/.style={purple, dashed}}

\pgfplotsset{fraction communicative chart/.style={
        ymin={0},
        ymax={1},
        ytick={0.0,0.5,1.0},
        grid,
        table/x=B0,
        table/y=communicative_fraction,
        cycle list={
          simulation scatter,
          full theory line,
          approx theory line,
        },
        enlarge y limits=0.05,
    }
}

\pgfplotsset{
  harmony/.style={fill=Paired-G,draw=Paired-G},
  chicken/.style={fill=Paired-F,draw=Paired-F}, 
  battle/.style={fill=Paired-E,draw=Paired-E}, 
  hero/.style={fill=Paired-H,draw=Paired-H}, 
  compromise/.style={fill=Paired-C,draw=Paired-C},
  concord/.style={fill=Paired-D,draw=Paired-D},
  staghunt/.style={fill=Paired-B,draw=Paired-B},
  dilemma/.style={fill=Paired-K,draw=Paired-K},
  deadlock/.style={fill=Paired-A,draw=Paired-A},
  assurance/.style={fill=Paired-J,draw=Paired-J},
  coordination/.style={fill=Paired-I,draw=Paired-I},
  peace/.style={fill=Paired-L,draw=Paired-L},
  neutral/.style={fill=gray,draw=gray},
  all_communicative/.style={fill=lightgray,draw=lightgray},
  all_noncommunicative/.style={fill=darkgray,draw=darkgray},
  disconnected_synchronized_populations/.style={fill=black,draw=black},
}

\pgfplotsset{colormap={strategycolors}%
{[1cm] color (0cm) = (blue!100) color (19cm) = (blue!05)
  color (20cm) = (red!100) color (39cm) = (red!05)}}

\pgfplotsset{table/col sep=comma}

\def\listterminator{;}
\makeatletter

\def\sortlist#1#2,#3#4,#5\relax{%
  \if\listterminator#3#4\relax%
    \edef\sortedlist{\sortedlist#1#2}%
  \else
    \ifnum\the\lccode`#1<\the\lccode`#3\relax%
      \edef\sortedlist{\sortedlist\presorted#1#2, }%
      \expandafter\def\expandafter\svfirst\expandafter{\presorted#3#4}%
      \def\presorted{}%
      \expandafter\sortlist\svfirst,#5\relax%
    \else%
      \ifnum`#1=`#3\relax%
        \ifx\relax#2\relax%
          \edef\sortedlist{\sortedlist\presorted#1#2, }%
          \expandafter\def\expandafter\svfirst\expandafter{\presorted#3#4}%
          \def\presorted{}%
          \expandafter\sortlist\svfirst,#5\relax%
        \else%
          \ifx\relax#4\relax%
            \edef\sortedlist{\sortedlist\presorted#3#4, }%
            \expandafter\def\expandafter\svfirst\expandafter{\presorted#1#2}%
            \def\presorted{}%
            \expandafter\sortlist\svfirst,#5\relax%
          \else
            \g@addto@macro\presorted{#1}%
            \sortlist#2,#4,#5\relax%
          \fi%
        \fi%
      \else%
        \let\tmp\sortedlist%
        \def\sortedlist{}%
        \expandafter\def\expandafter\svfirst\expandafter{\presorted#3#4}%
        \expandafter\def\expandafter\svsecond\expandafter{\presorted#1#2}%
        \def\presorted{}%
        \expandafter\expandafter\expandafter\expandafter\expandafter%
        \expandafter\expandafter\sortlist\expandafter\expandafter%
        \expandafter\tmp\expandafter\svfirst\expandafter,\svsecond,#5\relax%
      \fi%
    \fi%
  \fi%
}
\makeatother

\pgfplotsset{
    table/search path={../../data/processed},
}

\begin{document}

\title{Evolutionary Kuramoto dynamics unravels origins of chimera states
in neural populations}

\author{
\href{https://orcid.org/0000-0003-3039-172X}{Thomas Zdyrski}$^{1}$,
\href{https://orcid.org/0000-0002-0281-2868}{Scott Pauls}$^{1}$,
and
\href{https://orcid.org/0000-0001-8252-1990}{Feng Fu}$^{1,2}$
}

\affil{$^{1}$Department of Mathematics, Dartmouth College, Hanover, NH 03755\\
$^{2}$Department of Biomedical Data Science, Geisel School of Medicine at Dartmouth, Lebanon, NH 03756
}


\abstract{
Neural synchronization is central to cognition~\citep{santos2017chimera,bansal2019cognitive}.
However, incomplete synchronization often produces chimera
states~\citep{majhi2019chimera,hizanidis2016chimera},
where coherent and incoherent dynamics coexist.
While previous studies~\citep{abrams2004chimera}
have explored such patterns
using networks of coupled oscillators,
it remains unclear why neurons commit to communication
or how chimera states persist.
Here, we investigate the coevolution of neuronal phases
and communication strategies on directed, weighted networks,
where interaction payoffs depend on phase alignment~\citep{tripp2022evolutionary}
and may be asymmetric due to unilateral communication.
We find that both connection weights and directionality influence the stability
of communicative strategies---and, consequently,
full synchronization---as well as the strategic nature of neuronal interactions.
Applying our framework to the \glsfmtshort{celegans}
connectome~\citep{cook2019whole,towlson2013rich,yan2017network},
we show that emergent payoff structures, such as the snowdrift game,
underpin the formation of chimera states.
Our computational results demonstrate a promising neurogame-theoretic perspective,
leveraging evolutionary graph theory to shed light on mechanisms
of neuronal coordination beyond classical synchronization models.
}

\maketitle
\glsresetall

\section{Introduction}
Evolutionary game theory (EGT) is the application
of game theory to evolving populations
of individuals with behavioral strategies.
This tool is useful for studying how local interaction rules
yield large-scale patterns such as cooperation~\citep{sigmund1999evolutionary}
and has found use in fields including international politics, ecology,
and protein folding~\citep{traulsen2023future}.
Studies~\citep{cohen2009evolutionary,antonioni2017coevolution,tripp2022evolutionary}
have even applied EGT to non-reproducing neurons
by viewing neuron plasticity as an evolutionary process
where firing patterns change and are ``learned'' over time.
Evolutionary \emph{graph} theory places evolutionary games
on graphs to investigate the role of structure in population evolution.
Prior studies have found that structure
qualitatively changes game evolution.
For instance, cooperation is enhanced by
small-degree nodes~\citep{ohtsuki2006simple}
or unidirectional edges~\citep{su2022evolution}
and suppressed by weighted edges~\citep{bhaumik2024constant}.
Thus, the study of evolving, structured populations
should account for the effects of
incompleteness, directedness, and weightedness.

Kuramoto networks are groups of coupled oscillators
where the coupling strength depends sinusoidally
on the oscillators' phase difference.
These networks are popular
models for neuron behaviour~\citep{cabral2011role,deng2024chimera}
because they exhibit tunable synchronization.
Prior studies~\citep{antonioni2017coevolution}
have modelled interneuron communication
with Kuramoto oscillators and EGT
using the prisoner's dilemma game type.
Other studies~\citep{tripp2022evolutionary}
generalized this approach to include dynamically changing game types
by introducing an evolutionary Kuramoto (EK) model
to show how the relationship between communication benefit and cost
influences the emergence of synchronized communication
or non-communication regimes.

One intriguing aspect of neuron oscillations
is the observation of chimera states~\citep{majhi2019chimera}.
These states exhibit the simultaneous existence
of coherent and disordered phases~\citep{abrams2004chimera}.
Previous studies have proposed
chimera states as
a key component of human cognitive organization~\citep{bansal2019cognitive},
a facilitator of spiking and bursting phases~\citep{santos2017chimera},
and an outcome of modular networks~\citep{hizanidis2016chimera}.
Despite the observed importance of these chimera states,
the factors that give rise to coherent/disordered coexistence
remain incompletely characterized.

The nematode \gls{celegans}
is a model organism in neuroscience due to
its simple brain connectome~\citep{cook2019whole}
of only \num{302} neurons.
Despite their simplicity, models of the \gls{celegans} brain
still display a wide array of complex phenomena including
topologically-central rich clubs
crucial to motor neurons~\citep{towlson2013rich},
phenomenological connections to control theory~\citep{yan2017network},
and chimera states~\citep{hizanidis2016chimera}.

In this paper,
we introduce an asymmetric evolutionary Kuramoto model
and analyze its chimera-like states on the
\gls{celegans} connectome.
Our results connect individual neuron fitness and non-trivial brain topology
to chimera-like brain states.
Given current technological limitations with direct measurement of in-vivo neuron activity,
frameworks like ours create testable hypotheses connected to the theory's assumptions.
Our computational model represents a simple yet versatile framework
to illuminate the influence of neural connectivity on chimera-like brain states
beyond classical synchronization models.

\section{Results}\label{sec:results}

\subsection{Model}
\begin{figure*}
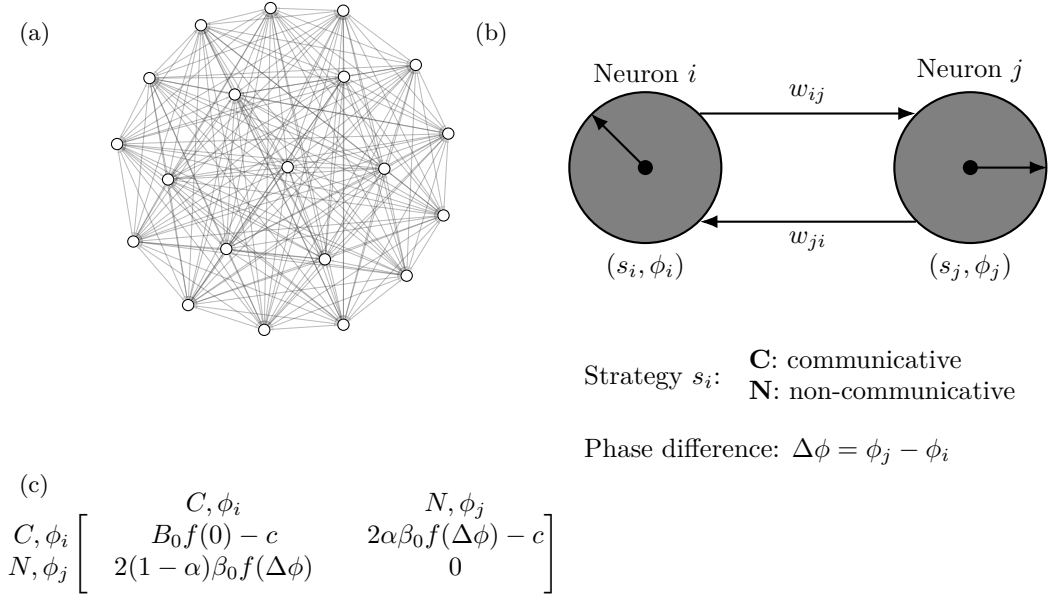

  \centering
  \begin{subcaptiongroup}
    \stackinset{l}{10pt}{t}{2.5in}%
      {\phantomcaption\label{fig:payoff_matrix}\captiontext*{}}{%
    \stackinset{l}{2.5in}{t}{10pt}%
      {\phantomcaption\label{fig:player_interactions}\captiontext*{}}{%
    \stackinset{l}{10pt}{t}{10pt}%
      {\phantomcaption\label{fig:graph_well-mixed}\captiontext*{}}{%
    {\includestandalone{tikz/model-setup}}%
  }}}
  \end{subcaptiongroup}
  \caption{
    \textbf{
      Evolutionary Kuramoto dynamics with weighted neural connectivity.
    }
    \protect{\subref{fig:graph_well-mixed}}
    The graph of a well-mixed population with $N=20$ players
    where each pair of players is connected by a directed edge in each direction.
    \protect{\subref{fig:player_interactions}}
    The connectivity between two sample players, $i$ and $j$,
    showing directed, weighted edges $w_{ij}$ and $w_{ji}$.
    Each player has a strategy (communicative $C$ or non-communicative $N$)
    and phase $\phi = 2\pi k/m$ with $k \in 0,\ldots,m-1$
    and $m$ the number of phases.
    \protect{\subref{fig:payoff_matrix}}
    The payoff matrix shows the reward the row-player $(C, \phi_i)$
    receives after playing a game with the column-player $(N, \phi_j)$
    assuming either player can switch strategy and phase to the other's.
  }\label{fig:connectivity}
\end{figure*}

We extend the EK model in two ways:
placing the player network on a directed, weighted graph;
and introducing a payoff asymmetry.
First, we represent a well-mixed population
as a complete graph in \cref{fig:graph_well-mixed}
with players represented by nodes and games by edges.
We generalize this to directed graphs
where game (edge) payoffs only flow to the head players (nodes).
We can also represent bidirectional games with
a pair of edges in both directions,
as shown in \cref{fig:player_interactions}.
For reference, the \gls{celegans} connectome
has \num{38} self-loops, \num{669} bidirectional edge pairs,
and \num{2331} unpaired edges.
Finally, we interpret the \gls{celegans} chemical connectome's
integer-valued edge weights
as the number of connections between nodes,
so these weights scale each payoff.

We also generalize the payoff structure to incorporate
an asymmetry between communicators and non-communicators.
We characterized each player (node) by its
strategy $s_i$---either communicative $C$ or non-communicative $N$---
and its phase $\phi_i$---taking one of $m = \num{20}$ evenly-spaced values
between $0$ and $2 \pi$.
When exactly one partner is communicative, the
asymmetry $\alpha \in [0,1]$ biases the payoff toward (against)
the communicator when $\alpha$ is greater (less) than $1/2$,
while $\alpha = 0.5$ reproduces the symmetric case.
\Cref{fig:payoff_matrix} shows the payoff matrix
for these mixed $CN$ or $NC$ interactions
and includes the maximum joint benefit $B_0 = 0.15$,
maximum mixed benefit $\beta_0$, cost $c$,
and sinusoidal Kuramoto coupling
$f(\Delta \phi) = [1+\cos(\phi_j - \phi_i)]/2$.

Unless otherwise specified,
each simulation uses
$m = 20$ phases,
selection strength of $\delta=0.2$,
mutation rate of $\mu=\num{1E-4}$,
cost $c$ of \num{0.1},
maximum joint benefit $B_0 = 0.15$,
maximum mixed benefit $\beta_0$ of $\num{0.95} B_0$,
and runs for \num{8E6} time steps
(even if only a smaller time series subset is shown).

\subsection{Parameter space}
\begin{figure*}
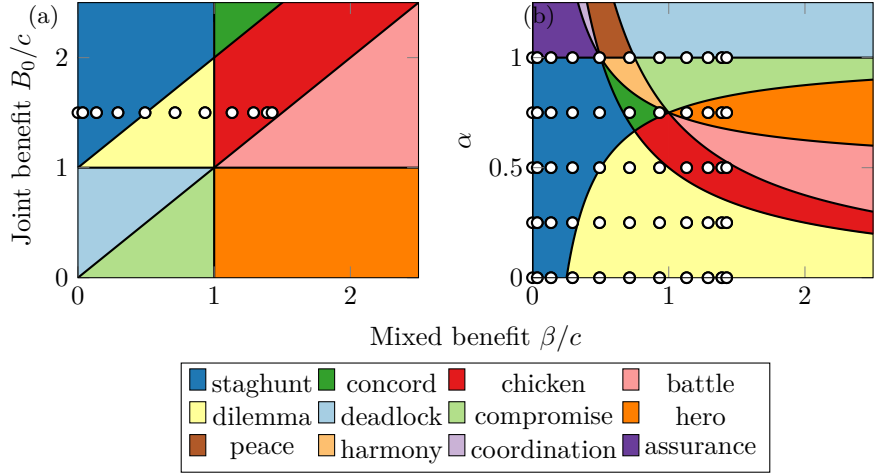

  \centering
  \begin{subcaptiongroup}
    \stackinset{l}{2.7in}{t}{0pt}%
      {\phantomcaption\label{fig:phase-diagram-beta_alpha}\captiontext*{}}{%
    \stackinset{l}{10pt}{t}{0pt}%
      {\phantomcaption\label{fig:phase-diagram-beta_B}\captiontext*{}}{%
    {\includestandalone{tikz/phase-diagram}}%
  }}
  \end{subcaptiongroup}
  \caption{
    \textbf{
      Payoff asymmetry enriches neural interactions
      well beyond the classic prisoner's dilemma game type.
      Region plots illustrate the diverse range
      of game types that neural populations can engage in
      during evolutionary dynamics.
    }
    Slices of the three-parameter mixed $CN$ game-type phase diagram
    in the
    \protect{\subref{fig:phase-diagram-beta_B}}
    $\beta$-$B_0$ plane ($\alpha = 0.5$)
    and
    \protect{\subref{fig:phase-diagram-beta_alpha}}
    $\beta$-$\alpha$ plane ($B_0/c = 1.5$).
    For two players with phase difference $\Delta \phi$,
    the mixed benefit is
    $\beta = \beta_0 [1 + \cos(\Delta \phi)]/2$.
    The legend displays the game type corresponding to each color.
    The white dots represent the $m=20$ potential phase differences
    as well as the restriction
    \protect{\subref{fig:phase-diagram-beta_B}}
    $B_0/c = 1.5$
    or
    \protect{\subref{fig:phase-diagram-beta_alpha}}
    $\alpha \in [0,0.25,0.5,0.75,1]$.
  }\label{fig:phase-diagram}
\end{figure*}

One of the key aspects of the EK model
is that multiple $2 \times 2$ game types can emerge
among the players during the population's evolution.
These game types~\citep{bruns2015names} include
dilemma (\aka{} ``prisoner's dilemma''),
deadlock (``anti-prisoner's dilemma''),
chicken (``hawk-dove'' or ``snowdrift''),
hero (``Bach or Stavinsky'' with the lowest two payoffs swapped),
harmony (game with strong incentive alignment),
or
concord (similar to harmony, weaker incentives).
See Section 2 of the supplementary material
for order graphs depicting the payoff structure of each game type.
While $CC$ interactions are always double-cooperation games
and $NN$ interactions are always neutral games,
mixed ($CN$ or $NC$) games show a variety of game types,
(\pcref{fig:phase-diagram}).
We can visualize these games
by looking at two-dimensional slices
of the three-dimension phase space
characterized by $\alpha$, $\beta \coloneqq \beta_0 f(\Delta \phi)$,
and $B_0$.
\cref{fig:phase-diagram-beta_B} shows a $\beta$-$B_0$ slice of phase space
and generalizes figure 1 of a prior study~\citep{tripp2022evolutionary},
while \cref{fig:phase-diagram-beta_alpha} shows a $\beta$-$\alpha$ slice.
Each straight line of white dots represent permissible values of the $m = 20$ phases
$\Delta \phi_i$ across our simulations
and highlight which regions of parameter space are accessible.

\subsection{Complete graphs}\label{sec:complete_graph}

\begin{figure*}
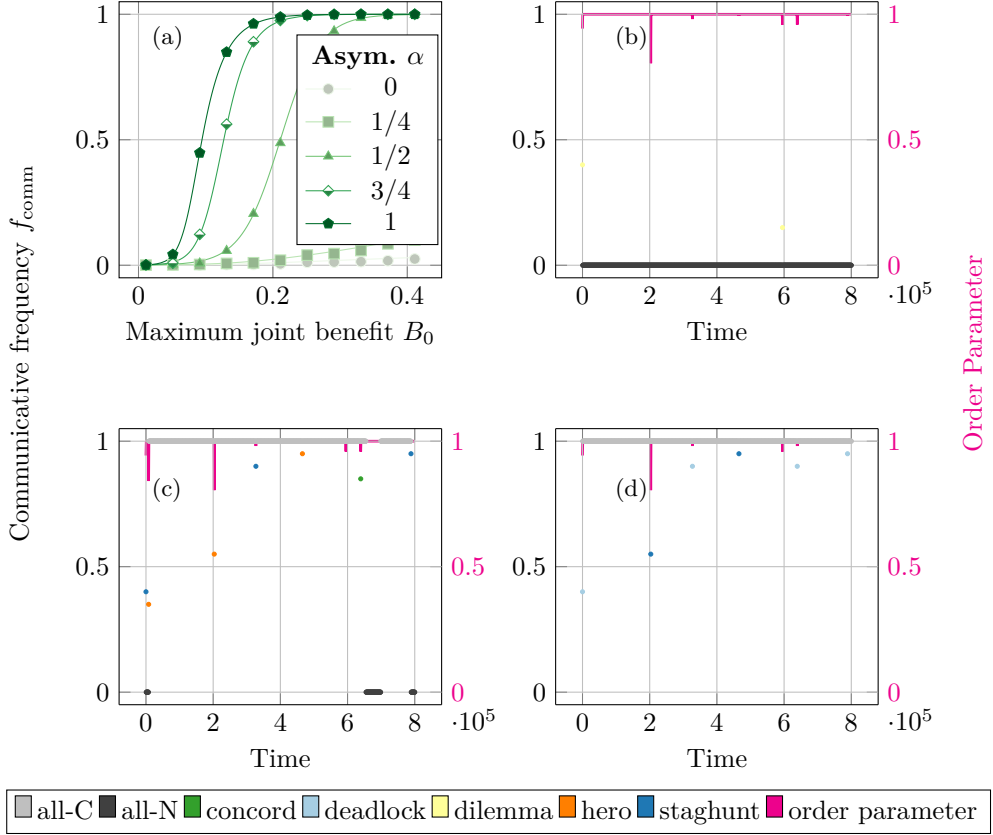

  \centering
  \begin{subcaptiongroup}
    \stackinset{l}{3.2in}{t}{2.5in}%
      {\phantomcaption\label{fig:time-series_well-mixed_alpha-1}\captiontext*{}}{%
    \stackinset{l}{0.8in}{t}{2.5in}%
      {\phantomcaption\label{fig:time-series_well-mixed_alpha-075}\captiontext*{}}{%
    \stackinset{l}{3.2in}{t}{10pt}%
      {\phantomcaption\label{fig:time-series_well-mixed_alpha-0}\captiontext*{}}{%
    \stackinset{l}{0.8in}{t}{10pt}%
      {\phantomcaption\label{fig:multi-comm-frac}\captiontext*{}}{%
    {\includestandalone{tikz/well-mixed}}%
  }}}}
  \end{subcaptiongroup}
  \caption{
    \textbf{
      Impact of symmetry breaking on neural synchronization
      in well-mixed populations.
    }
    Communication frequency $f_{\text{comm}}$ for the well-mixed topology.
    \protect{\subref{fig:multi-comm-frac}}
    Time-averaged $f_{\text{comm}}$ as a function
    of the maximum joint benefit $B_0$
    for different values of the payoff asymmetry $\alpha$.
    The marks represent the simulation results
    and the lines represent the theory predictions.
    \protect{\subref{fig:time-series_well-mixed_alpha-0}}--\protect{\subref{fig:time-series_well-mixed_alpha-1}}
    Scatter plots where
    the left vertical axes show the instantaneous $f_{\text{comm}}$
    as a function of time and are color-coded according
    to the plurality mixed game type as indicated in the legend.
    The right vertical axes give the order parameter $\rho$
    (\pcref{eq:order_parameter}), in magenta, as a function of time.
    The asymmetry is
    \protect{\subref{fig:time-series_well-mixed_alpha-0}}
    $\alpha = 0$,
    \protect{\subref{fig:time-series_well-mixed_alpha-075}}
    $\alpha = 0.75$,
    and
    \protect{\subref{fig:time-series_well-mixed_alpha-1}}
    $\alpha = 1$.
  }\label{fig:well-mixed}
\end{figure*}

First, we will explore the influence of the newly introduced asymmetry
on a $N=20$-player, well-mixed population.
\Cref{fig:multi-comm-frac} compares the frequency of communicative strategies
$f_{\text{comm}}$ to the maximum joint benefit $B_0$.
The marks represent the simulation results
and the lines represent the full analytic
result (Eq.\ 10 in supplementary material) for a well-mixed population.
The $B_0$ step size is \num{0.04},
and the simulations ran for \num{2E8} time steps.
In general, $f_{\text{comm}}$ is low for small $B_0$,
rises to $f_{\text{comm}} = 0.5$ at some break-even $B_0$,
and plateaus to $f_{\text{comm}} \approx 1$ for large $B_0$.
We can validate our model by comparing our
$\alpha = 0.5$ case with a previous study~\citep{tripp2022evolutionary}
to observe qualitatively similar results,
with our break-even $B_0 \approx 0.21$
corresponding to their $B_0 = 2 (N-1) c/(N-2) = 0.21$ break-even condition.
We also see that increasing (decreasing)
the asymmetry $\alpha$ dilates (stretches) this sigmoid function
in the $B_0$ direction.
This $\alpha$-dependence is reasonable,
as increasing $\alpha$ corresponds to biasing the payoff
in a mixed $CN$ interaction towards the communicative partner.

While \cref{fig:multi-comm-frac} displays
the time-averaged system state,
it is also useful to investigate the time-dependent variations.
\Crefrange{fig:time-series_well-mixed_alpha-0}{fig:time-series_well-mixed_alpha-1}
depict the frequency
of communicative strategies $f_{\text{comm}}$
as scatter plots of time on the left vertical axis
for different values of the asymmetry $\alpha$.
These time-series points are color-coded
grey if all players are communicative or
black if all players are non-communicative;
otherwise, the points are colored according
to the plurality mixed game type, as indicated in the legend.
On the right vertical axes,
magenta line plots depict the order parameter
$\rho \in [0,1]$
given by \cref{eq:order_parameter}.

\Cref{fig:time-series_well-mixed_alpha-0} shows the time-variation
when the system heavily favors non-communicative players
with $\alpha = 0$.
After an initial disordered, dilemma-type game (yellow),
the system synchronizes in the non-communicative regime
(black line at $f_{\text{comm}} = 0$)
with only a brief dilemma-type game excursion near time step \num{6e5}.
\Cref{fig:time-series_well-mixed_alpha-075} depicts an asymmetry $\alpha = 0.75$
that moderately encourages communicativeness.
The system is synchronized in a communicative state
(gray line at $f_{\text{comm}} = 1$) \SI{92}{\percent} of the time,
with unstable excursions to synchronized, non-communicative states
(black line at $f_{\text{comm}} = 0$)
and disordered, hero (orange) and staghunt (dark blue) game types.
This \SI{92}{\percent} communicative rate is higher than the expected \SI{75}{\percent}
from \cref{fig:multi-comm-frac} (for $B_0 = 0.15$ and $\alpha = 0.75$),
likely due to the small timespan shown in
\cref{fig:time-series_well-mixed_alpha-075}.
Finally, \cref{fig:time-series_well-mixed_alpha-1} shows the
$\alpha = 1$ case where communication is heavily incentivized.
Like the \subref{fig:time-series_well-mixed_alpha-0} $\alpha = 0$ case,
the system is virtually always synchronized in the communicative regime,
with only brief excursions to deadlock (light blue)
or staghunt (dark blue) games types.
We note that all three
\subref{fig:time-series_well-mixed_alpha-0}-\subref{fig:time-series_well-mixed_alpha-1}
scenarios are virtually always synchronized;
the order parameter $\rho$ (magenta) is almost always $\rho = 1$,
with only occasional dips.
The temporary drops in communicative frequency $f_{\text{comm}}$
and order parameter $\rho$ are likely the result of mutations
which occur, on average, every $1/\mu = \num{1E4}$ time steps.

\subsection{\glsfmtshort{celegans} graphs}\label{sec:elegans_graph}

Here, we consider the weighted, directed hermaphroditic \gls{celegans}
chemical connectome~\citep{cook2019whole}.
For reference, \cref{fig:graph} shows the $N = 300$ nodes and the directed edges
(we exclude the two unconnected neurons CANL/CANR).
The nodes are colored according to their strategy at a particular time step,
with blues representing communicative strategies
and reds representing non-communicative ones,
and different shades corresponding to different phases $\phi$.
We note the chimera-like character
of the large, synchronized group of red nodes
coexisting with disordered neighboring nodes.

\begin{figure*}
  \centering
  \begin{subcaptiongroup}
    \stackinset{l}{3in}{t}{3.3in}%
      {\phantomcaption\label{fig:time-series_celegans-full}\captiontext*{}}{%
    \stackinset{l}{3in}{t}{0.4in}%
      {\phantomcaption\label{fig:graph}\captiontext*{}}{%
    \stackinset{l}{0.65in}{t}{4.1in}%
      {\phantomcaption\label{fig:comm-frac_celegans-undirected}\captiontext*{}}{%
    \stackinset{l}{0.65in}{t}{2.1in}%
      {\phantomcaption\label{fig:comm-frac_celegans-unweighted}\captiontext*{}}{%
    \stackinset{l}{0.65in}{t}{0.1in}%
      {\phantomcaption\label{fig:comm-frac_celegans-full}\captiontext*{}}{%
    {\includestandalone{tikz/c-elegans}}%
  }}}}}
  \end{subcaptiongroup}
  \caption{
    \textbf{
      Rise of chimera states in \gls{celegans} neural network.
    }
    \protect{\subref{fig:comm-frac_celegans-full}}--\protect{\subref{fig:comm-frac_celegans-undirected}}
    Time-averaged fraction of players that are communicative as a function
    of the maximum joint benefit $B_0$.
    The network topologies are the
    \protect{\subref{fig:comm-frac_celegans-full}}
    weighted, directed \gls{celegans} connectome,
    \protect{\subref{fig:comm-frac_celegans-unweighted}}
    unweighted, directed \gls{celegans} connectome,
    and
    \protect{\subref{fig:comm-frac_celegans-undirected}}
    weighted, undirected \gls{celegans} connectome.
    \protect{\subref{fig:graph}}
    The network topology for the \gls{celegans}
    $N=300$ weighted, directed connectome
    with asymmetry $\alpha = 0.75$.
    The colors represent the $2m = 40$ strategies
    at a particular time step;
    blue colors are communicative,
    red colors are non-communicative,
    and shades represent different phases $\phi$.
    \protect{\subref{fig:time-series_celegans-full}}
    Scatter plot with the same axes and coloring
    as \protect{\crefrange{fig:time-series_well-mixed_alpha-0}{fig:time-series_well-mixed_alpha-1}}
    showing the communicative frequency, plurality mixed game-types, and order parameter.
  }\label{fig:c-elegans}
\end{figure*}

Next, to quantify the observations of \cref{fig:graph},
\cref{fig:comm-frac_celegans-full} shows the fraction of players
using communicative strategies $f_{\text{comm}}$ averaged across
the entire simulation of \num{2E8} time steps as a function
of the maximum joint benefit $B_0$ in \num{0.04} steps
for the \gls{celegans} connectome network topologies.
All subplots use an asymmetry of $\alpha=0.75$.

We note that the behaviour
of the \subref{fig:comm-frac_celegans-full} \gls{celegans} case
is qualitatively distinct
from the $\alpha = 0.75$ well-mixed case in \cref{fig:multi-comm-frac}.
The communicative fraction $f_{\text{comm}}$
is flatter for $B_0 \le 0.08$,
has a steep jump to \num{0.77} at $B_0 = 1.6$,
and decreases to a horizontal asymptote around \num{0.60}.
We can isolate the cause of this deviation
from the \cref{fig:multi-comm-frac} well-mixed
behavior by looking at variations to the \cref{fig:comm-frac_celegans-full}
\gls{celegans} network topology.
First, the \subref{fig:comm-frac_celegans-unweighted}
\emph{directed}, unweighted connectome
is qualitatively similar to the
\subref{fig:multi-comm-frac}
well-mixed case with a monotonic increase
from low communicativeness for $B_0 < 0.1$
to full communicativeness for $B_0 \ge 0.2$.
This implies that directedness
plays only a small role in the qualitative shape
of the \subref{fig:comm-frac_celegans-full} full \gls{celegans} case.
Conversely,
the \subref{fig:comm-frac_celegans-undirected}
\emph{weighted}, undirected connectome
looks similar to the \subref{fig:comm-frac_celegans-full}
full \gls{celegans} case,
displaying the same plateau at $f_{\text{comm}} \approx 0.7$,
though the peak around $B_0 = 0.15$ is less pronounced.
This similarity implies that the connectome's edge weights
cause most of the deviation
between the \subref{fig:comm-frac_celegans-full} full \gls{celegans}
case and the \cref{fig:multi-comm-frac} well-mixed case.

We can also investigate the time-evolution of the \gls{celegans} system
using the same parameters as the
\crefrange{fig:time-series_well-mixed_alpha-0}{fig:time-series_well-mixed_alpha-1}
well-mixed case but with the \gls{celegans} connectome graph.
Compared to the well-mixed case,
the \subref{fig:time-series_celegans-full} \gls{celegans} case
depicts a far more heterogeneous population.
Here, the population never stabilizes
to a fully communicative or non-communicative state.
Instead, both its communicative frequency and order parameter
stay between \SIrange{40}{70}{\percent}.
The mean communicative frequency $f_{\text{comm}} \approx \SI{60}{\percent}$
is similar to the expected $f_{\text{comm}} \approx \SI{70}{\percent}$
from \cref{fig:comm-frac_celegans-full} with $B_0 = 0.15$;
the discrepancy likely comes from the stochasticity
in this small, \num{8E5} time-step subset.
And while the \cref{fig:time-series_well-mixed_alpha-075} well-mixed setup
displays hero, staghunt, and concord plurality mixed game-types,
the \cref{fig:time-series_celegans-full} \gls{celegans} setup
only displays hero and staghunt games.
Finally, the \gls{celegans} case is less stable to mutations
than the \cref{fig:time-series_well-mixed_alpha-075} well-mixed case:
instead of stable synchronized states with transient impulses,
the \gls{celegans} case depicts disordered states with discontinuous offsets.

\subsection{Chimera-like index}
\begin{figure*}
  \centering
  \begin{subcaptiongroup}
    \stackinset{l}{3in}{t}{2.5in}%
      {\phantomcaption\label{fig:graph_celegans_asymmetry1}\captiontext*{}}{%
    \stackinset{l}{0.71in}{t}{2.5in}%
      {\phantomcaption\label{fig:graph_celegans_asymmetry0}\captiontext*{}}{%
    \stackinset{l}{3in}{t}{15pt}%
      {\phantomcaption\label{fig:metastability_index}\captiontext*{}}{%
    \stackinset{l}{0.7in}{t}{15pt}%
      {\phantomcaption\label{fig:chimera_index}\captiontext*{}}{%
    {\includestandalone{tikz/chimera-states}}%
  }}}}
  \end{subcaptiongroup}
  \caption{
    \textbf{
      Characterizing chimera states.
    }
    \protect{\subref{fig:chimera_index}}
    The chimera-like index (\protect{\pcref{eq:chimera_index}})
    and
    \protect{\subref{fig:metastability_index}}
    metastability index (\protect{\pcref{eq:metastability_index}})
    as functions of the asymmetry $\alpha$
    for the
    weighted, directed \gls{celegans} connectome.
    To calculate the indices,
    we split the graph into two communities according to the nodes'
    relative covariance (\cf{} \cref{sec:chimera-metastability-def}).
    \protect{\subref{fig:graph_celegans_asymmetry0}}--\protect{\subref{fig:graph_celegans_asymmetry1}}
    A snapshot of the \gls{celegans} connectome where
    blue colors are communicative,
    red colors are non-communicative,
    and shades represent different phases $\phi$.
    The asymmetry is
    \protect{\subref{fig:graph_celegans_asymmetry0}}
    $\alpha = 0$
    or
    \protect{\subref{fig:graph_celegans_asymmetry1}}
    $\alpha = 1$.
  }\label{fig:chimera-index}
\end{figure*}

Time-lapse animations of the system's time evolution
show a subset of the nodes exhibiting high synchronicity
with others remaining disordered: this is characteristic of a chimera state.
Animations depicting this phenomena are available as supplementary videos,
and an interactive website for exploring the full data set
through plots and animations is provided in the ``Code availability''
section at the end of the paper.
In order to quantify this chimera-like effect, we will investigate
the chimera-like index $\chi$ (\pcref{eq:chimera_index})
and metastability index $\lambda$ (\pcref{eq:metastability_index}).
As discussed in \cref{sec:chimera-metastability-def},
the chimera-like index
measures the coherence difference between communities of players,
is equal to the time-averaged community covariance,
and has a theoretical maximum value
of $M/[4(M-1)] = \num{0.5}$~\citep{shanahan2010metastable}
for our $M=2$ communities.
Conversely, the metastability index
represents how often the system transitions
between synchronicity and disorder,
is equal to the community-average of the temporal covariance,
and has a practical maximum of \num{0.08}~\citep{shanahan2010metastable}.
\Cref{fig:chimera-index} shows the
\subref{fig:chimera_index} chimera-like index $\chi$
and
\subref{fig:metastability_index} metastability index $\lambda$
as functions of the asymmetry $\alpha$.
These simulations ran with the same parameters as the
\crefrange{fig:comm-frac_celegans-full}{fig:comm-frac_celegans-undirected}
\gls{celegans} time-series data.

The metastability displayed in \cref{fig:chimera-index}
is less than \num{0.01}, much smaller
than the practical maximum of \num{0.08}.
This implies that the system has low metastability
and spends most of its time at a nearly constant
synchronicity $\rho_m(t)$.
Furthermore, while the metastability is higher (more stability variations)
at $\lambda \approx \num{0.01}$ for the $\alpha = 0$ case,
the metastability drops precipitously for larger $\alpha$,
always staying below \num{0.004} implying that
higher asymmetry $\alpha$ makes the system more stable.
Conversely, the chimera-like index $\chi$
indicates a high chimeric quality
with values (\numrange{0.06}{0.13})
nearly a quarter of the theoretical maximum (\num{0.5}).
Given that we average the chimera-like index over time,
the system's deviation from a complete chimera state
arises from both imperfect separation of the coherent/disordered populations
as well as time fluctuations in the chimeric quality.
While we observe that $\chi$ has a maximum at $\alpha = \num{0.75}$,
we caution that this is likely the result of calculating the communities
based on the $\alpha = \num{0.75}$ covariances.
Nevertheless, the high $\chi$ value for all asymmetries $\alpha$
implies that the strong chimeric character is intrinsic
rather than an artifact of our $\alpha = \num{0.75}$ choice.

\subsection{Game types}
\begin{figure*}
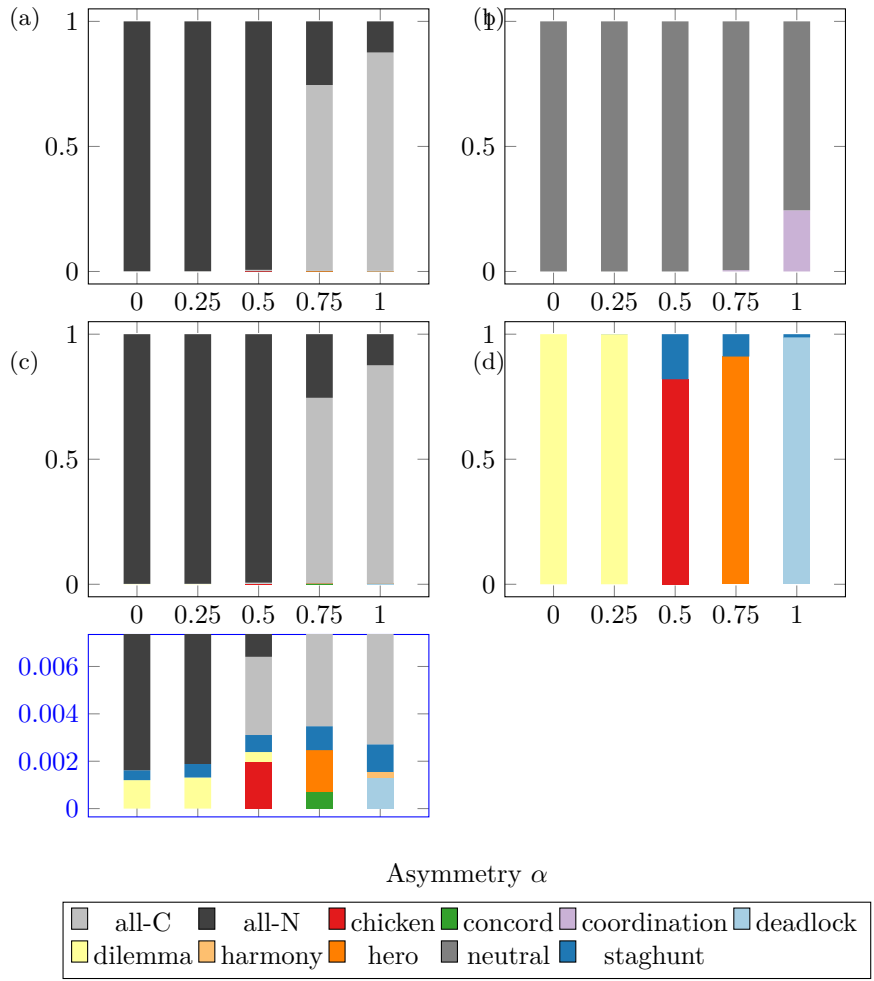

  \centering
  \begin{subcaptiongroup}
    \stackinset{l}{2.4in}{t}{1.8in}%
      {\phantomcaption\label{fig:game-type_celegans-full-only-mixed-games}\captiontext*{}}{%
    \stackinset{l}{0in}{t}{1.8in}%
      {\phantomcaption\label{fig:game-type_well-mixed-only-mixed-games}\captiontext*{}}{%
    \stackinset{l}{2.4in}{t}{0.0in}%
      {\phantomcaption\label{fig:game-type_celegans-full}\captiontext*{}}{%
    \stackinset{l}{0in}{t}{0.0in}%
      {\phantomcaption\label{fig:game-type_well-mixed}\captiontext*{}}{%
    {\includestandalone{tikz/game-types}}%
  }}}}
  \end{subcaptiongroup}
  \caption{
    \textbf{
      Influence of payoff asymmetry on game types played.
    }
    The plurality game type amongst all player interactions
    expressed as a fraction of all games played
    for different values of the asymmetry $\alpha$.
    The game types are color-coded according to the legend;
    additionally, ``all-C'' and ``all-N'' represent when the population
    is entirely synchronized to communicativeness or non-communicativeness, respectively.
    The network topologies are the
    \protect{\subref{fig:game-type_well-mixed}}
    $N=20$ well-mixed population
    and
    \protect{\subref{fig:game-type_celegans-full}}
    weighted, directed \gls{celegans} connectome.
    Similarly,
    \protect{\subref{fig:game-type_well-mixed-only-mixed-games}},
    \protect{\subref{fig:game-type_celegans-full-only-mixed-games}}
    depict the plurality game type when only considering mixed game types
    for the
    \protect{\subref{fig:game-type_well-mixed-only-mixed-games}}
    $N=20$ well-mixed population
    and
    \protect{\subref{fig:game-type_celegans-full-only-mixed-games}}
    weighted, directed \gls{celegans} connectome.
    A blue inset below the
    \protect{\subref{fig:game-type_well-mixed-only-mixed-games}}
    well-mixed panel shows the magnified lower region.
  }\label{fig:game-type}
\end{figure*}

Next, we seek to understand the types of games
that nodes play during the population's evolution.
In \cref{fig:game-type}, we investigate the plurality game type
amongst all player interactions at each time step~(\cf{} \cref{sec:plurality_game_type}).
\Cref{fig:game-type} shows the fraction of time
that a given game type is the plurality for different values
of the asymmetry $\alpha$ for
\subref{fig:game-type_well-mixed}
$N=20$ well-mixed,
and
\subref{fig:game-type_celegans-full}
\gls{celegans} connectome
network topologies.
The \subref{fig:game-type_well-mixed} well-mixed case
fully-synchronized to all-communicative or all-noncommunicative
for over \SI{99.6}{\percent} of the runtime
across every asymmetry $\alpha$,
as corroborated by the order parameter $\rho \approx 1$ in
\crefrange{fig:time-series_well-mixed_alpha-0}{fig:time-series_well-mixed_alpha-1}.
Furthermore,
the $\alpha$-dependence of the
communicative synchronization (``all-C'') fraction
approximates the communicative frequency $f_{\text{comm}}$
in \cref{fig:multi-comm-frac} for $B_0=0.15$.
In contrast, the
\subref{fig:game-type_celegans-full} \gls{celegans} system
is never synchronized, but instead is virtually always dominated by
coordination ($CC$) and neutral ($NN$) game types.

In order to investigate the other game types involved,
we can also consider the plurality \emph{mixed} game type
(\cf{} \cref{sec:plurality_game_type}).
The \subref{fig:game-type_well-mixed-only-mixed-games} well-mixed case
only changes for the $< \SI{1}{\percent}$ of the time when unsynchronized.
In order to better observe these games types,
a blue inset below \cref{fig:game-type_well-mixed-only-mixed-games}
magnifies these small fractions.
In contrast, the \subref{fig:game-type_celegans-full-only-mixed-games}
shows a variety of most-frequent game types
that vary based on the asymmetry:
dilemma for $\alpha = 0, 0.25$, chicken for $\alpha = 0.5$,
hero for $\alpha = 0.75$, and deadlock for $\alpha = 1$.
We note that these dominant game types
are the same as the most-frequent \emph{disordered}
(\ie{} states other than ``all-C'' or ``all-N'')
plurality mixed games as those played in
the \cref{fig:game-type_well-mixed-only-mixed-games}
well-mixed setup (\cf{} blue inset) for each $\alpha$.
Additionally, the \gls{celegans} system displays
some game-type heterogeneity, with
a second game type (staghunt) being the plurality
\SIrange{1}{18}{\percent} of the time
for asymmetries $\alpha \ge 0.5$.
Overall, we see that graph structure decreases synchronization
and asymmetry influences the dominant game types, which, in turn,
underpin the formation of chimera states.

\section{Discussion}\label{sec:discussion}

We can validate our \cref{fig:game-type_well-mixed-only-mixed-games} well-mixed
time-series results by comparing the $\alpha = 0.5$ case
to previous studies~\citep{tripp2022evolutionary}.
Their weak-selection $\delta = 0.2$
results also show the system spending
virtually all of its time in a synchronized state.
When not synchronized, their system had plurality
chicken (denoted ``snowdrift'' therein)
and staghunt (denoted ``coordination'',
not to be confused with the coordination-type game here) games.
Indeed, \cref{fig:phase-diagram-beta_B}
shows that staghunt, dilemma, and chicken
would be the only game types accessible for $\alpha = 0.5$,
which is corroborated in our simulations by the inset below
\cref{fig:game-type_well-mixed-only-mixed-games}.

Next, we will discuss the observed chimera states.
While the chimera-like quality is inherently time-dependent
(\cf{} \pcref{eq:chimera_index})
carefully chosen snapshots can still convey some of the chimera-like aspects.
\Cref{fig:graph_celegans_asymmetry0} shows a snapshot of the communication strategies
for asymmetry $\alpha = 0$ using the same color scheme as \cref{fig:graph}.
We notice a large region of light red representing
a synchronized group of non-communicators,
as well as a mix of disordered neighbors.
This coexistence of strong synchronization and disorder
is consistent with the high
chimera-like index $\chi$ observed in \cref{fig:chimera-index} for $\alpha = 0$.
Likewise, the $\alpha = 0.75$ case depicted in \cref{fig:graph}
has an even higher $\chi = \num{0.13}$,
consistent with the coexistence of a large,
synchronized non-communicative group (red)
and disordered neighbors (varying phases/colors).
In contrast,
\cref{fig:graph_celegans_asymmetry1} shows a snapshot for $\alpha = 1$,
for which we expect a lower chimera-like index of $\chi = \num{0.06}$.
Indeed, we see that the large communicative (red) group
in \cref{fig:graph_celegans_asymmetry1}
or \cref{fig:graph} has fractured into two
communicative (blue) groups with different phases (color shade),
lowering the chimeric quality.
Therefore, these snapshots give a glimpse of the chimera-like quality
in the \gls{celegans} populations.

We also note that \cref{fig:game-type} shows that
the high chimera-like index setups ($\alpha = 0.5, 0.75$)
are dominated by game types (chicken, hero, and battle),
shown in warm colors (red, orange, and pink, respectively),
with exponentially slow fixation times~\citep{antal2006fixation}.
In contrast, the lower chimera-like index setup ($\alpha = 0,0.25,1$)
is dominated by non-exponentially-slow dilemma and deadlock type games.
This connection between exponentially-slow fixation time games and high chimera index
seems suggestive and requires further investigation.

Another study of the \gls{celegans} connectome also found
chimera-like states using an entirely non-game-theory tool,
modular neural networks~\citep{hizanidis2016chimera}.
That study found a maximum chimera-like index of \num{0.12},
in agreement with our value of \num{0.13}, despite the different network models
of the \gls{celegans} connectome considered.
This agreement between two different models of neurons
reveals that the emergence of chimera states
is most likely primarily driven by the connectivity structure itself.

Chimera states of networked oscillators
exhibit coexisting
synchronized and disordered populations,
are present
in brains~\citep{santos2017chimera,bansal2019cognitive},
and may be pivotal in human cognition~\citep{bansal2019cognitive}.
Prior studies~\citep{deng2024chimera} investigated
chimera-like brain states in \gls{dmelanogaster}
using sinusoidally-coupled Kuramoto oscillators,
a frequent model for neuron dynamics~\citep{cabral2011role}.
However, the small-scale evolutionary factors
leading to chimera-like states remains poorly understood.
In this paper, we extended the evolutionary Kuramoto model
to include weighted, structured interaction graphs
and an asymmetry between
the communicator and non-communicator payoffs.
We first applied this model to a well-mixed population
and found that increasing (decreasing) the asymmetry
inhibits (promotes) communicativity.
Next, we applied the model to
a family of graphs deriving from the \gls{celegans} connectome.
This revealed that the graph's weightedness
has a much stronger influence on the communication frequency
than the graph's directedness does.
Finally, while the well-mixed players have homogeneous populations
that occasionally switch between synchronized communicative
and  non-communicative states,
the \gls{celegans} population remains far more heterogenous
with a stable, chimeric mix of disordered game types.

Future work will focus on applying the EK model
to families of generative brain network models~\citep{betzel2016generative}.
The limited number of empirical graphs analyzed in this study
hinders the identification of exact relationships
between the communicative fraction and graph properties
such as edge degree, weight, and directionality.
However, applying the model to parameterized families of graphs could allow
for fine-tuning the parameters to extract these relations.
Additionally, since this study only considers a single species' connectome,
it is unclear if our findings regarding the dependence
of chimeric activity on graph structure are generalizable.
Therefore, it is of interest to investigate this connection further
by applying our EK setup to other model organisms,
such as that of
\glsxtrlong{dmelanogaster}~\citep{schlegel2024whole}.

A key takeaway from this study was the importance
of edge weight on communicativity;
this has important parallels to neural computing,
where edge weights are a primary driver of functionality.
Additionally, the observation of chimeric states
arising from such simple neurogame-theoretic models
implies that the nature of neuron interactions is likely
a key component in producing these critical brain states.
Overall, evolutionary graph theory allowed us
to connect low-level payoff details for individual neurons
to high-level phenomena such as chimera-states,
and this model could serve as a valuable computational framework
for clarifying the influence of network structure on neural dynamics.

\section{Methods}\label{sec:methods}

\subsection{Game setup}\label{sec:game_setup}
We model the system of evolving, coupled oscillators
by discretizing the $2\pi$ phase angle into $m$ discrete phases $2 \pi j/m$
for $j \in 0, \ldots, m-1$.
The game's strategy space is the outer product of the $m$ phases
($m = 20$ for our simulations) and $2$ communicative choices,
communicative $C$ and non-communicative $N$.
For a given pair of phases, $\phi_i$ and $\phi_j$, the game is specified
by the payoff matrix.
If one player is communicative and the other is non-communicative
(``mixed game'', $CN$ or $NC$), the payoff matrix is
\begin{equation}
\begin{bNiceMatrix}[first-col,first-row,hvlines]
  & C, \phi_i & N, \phi_j \\
  C, \phi_i & B_0 f(0) - c & 2 \alpha \beta_0 f(\Delta \phi) - c \\
  N, \phi_j & 2 (1 - \alpha) \beta_0 f(\Delta \phi) & 0
\end{bNiceMatrix},
\label{eq:payoff-matrix}
\end{equation}
if both players are communicative ($CC$), the matrix is
\begin{equation}
\begin{bNiceMatrix}[first-col,first-row,hvlines]
  & C, \phi_i & C, \phi_j \\
  C, \phi_i & B_0 f(0) - c & B_0 f(\Delta \phi) - c \\
  C, \phi_j & B_0 f(\Delta \phi) - c & B_0 f(0) - c
\end{bNiceMatrix},
\end{equation}
and if both are non-communicative ($NN$), the matrix is
\begin{equation}
\begin{bNiceMatrix}[first-col,first-row,hvlines]
  & N, \phi_i & N, \phi_j \\
  N, \phi_i & 0 & 0 \\
  N, \phi_j & 0 & 0
\end{bNiceMatrix},
\end{equation}
with $f(\Delta \phi) = [1+\cos(\phi_j - \phi_i)]/2$
(\pcref{fig:payoff_matrix}).
Here, $B_0$, $\beta_0$, $\alpha$, and $c$ are fixed parameters
defining the game.
The cost $c$ represents the penalty paid by communicative players,
and $B_0$ and $\beta_0$ are the maximum benefits paid with
joint $CC$ communicators and mixed $CN$ players, respectively.
The phase-dependent function $f(\Delta \phi)$ encodes
the influence of phase mismatches.
Finally, the benefit asymmetry $\alpha \in [0,1]$ breaks the symmetry
between the payoff for the communicator and the non-communicator
when exactly one player is communicative.

\subsection{Population setup}\label{sec:pop_setup}

Given $N$ players, we associate a pair
of weighted, directed graphs to the population.
First, we use an interaction graph with $N$ nodes representing players
and weighted, directed edges representing games between players.
Second, we implement a reproduction graph with the $N$ nodes
still representing players
but the edges now corresponding to the ability of nodes to replace one another.
For simplicity, our reproduction graphs are identical to the interaction graph
with a single self-loop added to each node.
These self-loops are necessary in the reproduction graph
to ensure that each node has positive indegree
as required by the Moran process described in the next section.

\subsection{Birth-death Moran process}\label{sec:evo_setup}
The population is updated according to a birth-death Moran process
with exponential fitness~\citep{lieberman2005evolutionary}.
On each turn, the following steps are performed.
First, each edge in the interaction graph corresponds to a game
between head node $i$ and tail node $j$,
the edge's payoff $\pi_{ji}$ is scaled by the edge weight $w_{ji}$,
and the relevant payoff $w_{ji} \pi_{ji}$ is awarded to the head node only.
Since the \gls{celegans} edge weights are integers,
we can also interpret each edge weight as the number of games played
between the two nodes.
\Cref{fig:player_interactions} shows an illustration of this process
for a single pair of interacting players connected by a pair of
weighted, directed edges.
The total fitness for node $i$ is the exponential of the product
between the selection strength $\delta$
and the sum of payoffs to node $i$,
or $f_i = \exp(\delta \sum_j w_{ji} \pi_{ji})$ with the sum
over all edges inwardly incident to node $i$.
Then, a single focal node is chosen for reproduction
with probability proportional to the node's fitness $f_i$.
Finally, a node is chosen for replacement amongst the birth node's out-neighbors
with probability proportional to the reproduction graph's edge weight.
With mutation probability $\mu$,
the death node is replaced by a player with a uniformly random strategy;
otherwise, it is replaced by a player with the same strategy as the birth node.
This birth-death process is repeated for each turn.

\subsection{Communication frequency}
We define the frequency of communicative strategies
$f_{\text{comm}}(t)$ as the fraction of players employing
a communicative strategy $C$ at a given time step.
We also define the time-averaged communicative frequency
$f_{\text{comm}}$
by averaging $f_{\text{comm}}(t)$ over the entire simulation.
For simulation times long compared to the mutation turnover time
$T_{\text{turn}} \gg N/\mu$,
the initial, random distribution of strategies will be negligible
and the time-average will correspond to the long-time limit.
In section 2 of the supplementary material,
we derive an analytic expression (Eq.\ 10 of supplementary materials) for
$f_{\text{comm}}$ in the well-mixed case
by incorporating the benefit asymmetry $\alpha$.

\subsection{Game type nomenclature}
\label{sec:game-type-nomenclature}
Every edge of the interaction graph
defines a game between the two players it connects.
Using their relative phase difference $\Delta \phi$,
we can calculate the payoff matrix.
By comparing the order of each of the four entries,
we determine the game type using
a topological taxonomy~\citep{bruns2015names}.
Using this taxonomy, we calculate the ordinal rank of the four entries
in the \cref{eq:payoff-matrix} payoff matrix
and assign a unique name (\eg{} dilemma, deadlock, chicken, \etc{})
to each strict, symmetric game type.
However, \cref{fig:phase-diagram-beta_alpha}
shows that some mixed $CN$ games lie on the border between two game types,
such as when $\alpha = 1$ or $B/c = 0$
(bordering game types for $B/c > 0$ and $B/c < 0$, not shown).
The taxonomy~\citep{bruns2015names} classifies these non-strict games
according the number and location of ties (``high tie'', ``middle tie'',
``double tie'', \etc{}).
It also defines a convention for choosing one of the neighboring game types
to get a binomial nomenclature (\eg{} ``high harmony'', ``mid compromise'',
``double coordination'', \etc{}).
We follow the same convention for choosing a neighboring game
but drop the tie-indicator to keep our figure legends simple.
Specifically, referring to \cref{fig:phase-diagram-beta_alpha},
the $\alpha = 1$ tie between deadlock and compromise games
formally corresponds to ``low [dead]lock'',
but we label it as deadlock;
similarly, the tie between assurance and staghunt
is ``mid staghunt'', but we denote it as staghunt.
Likewise, all of the $B/c = 0$ payoff matrices
correspond to the ``double harmony'' game type,
but we denote them as simply ``harmony''.
Finally, the $NN$ game type is always ``neutral''
while the $CC$ game type is always ``double cooperation'',
which we denote as just ``cooperation''.

\subsection{Plurality game type}\label{sec:plurality_game_type}
At each time step, we calculate the game type for each edge
of the interaction graph.
We determine the game type by creating a two-by-two payoff matrix of possibilities
where both players have the hypothetical option of switching to the other player's
strategy/phase pairing.
Using the taxonomy discussed in \cref{sec:game-type-nomenclature},
we then assign a game type to that interaction.
We then identify the plurality game type across all player interactions,
where each edge's count is weighted by its edge weight.
Then, we calculate the frequency of this ``plurality game'' across
all time steps of a given simulation to determine the distribution
of games commonly played.
This ``plurality game-type'' metric, as depicted in
\cref{fig:game-type_well-mixed} and \cref{fig:game-type_well-mixed},
is often dominated by neutral games (between $NN$ players)
and coordination games (between $CC$) players.
To isolate the other game types involved,
we also defined a ``plurality \emph{mixed} game-type''
by only counting games between mixed $CN$ or $NC$ pairs in the plurality
(or labelling the time step as ``all-communicative''/``all-noncommunicative'',
as necessary).
This ``plurality mixed game-type'' shows more variety
and is depicted in all of the other figures
(\pcref{fig:well-mixed}, \pcref{fig:c-elegans},
\cref{fig:game-type_well-mixed-only-mixed-games}
and \cref{fig:game-type_well-mixed-only-mixed-games}
).
Note that an edge's game type is dependent
on the players' dynmically-evolving communication strategies (N or C)
and relative phase $\Delta \phi$,
as well as the fixed game parameters $c$, $B_0$, $\beta_0$, and $\alpha$.
Furthermore, this metric is only sensitive to the
plurality game and therefore provides no information
on the presence/absence of minority game types.

\subsection{Order parameter}
Given that the Kuramoto system of coupled oscillators
inspired this evolutionary game model,
we also define the standard Kuramoto order parameter:
\begin{equation}
  \rho = \frac{1}{N} \abs{\sum_{j=1}^N e^{\im \phi_j}}
  \label{eq:order_parameter}
\end{equation}
This parameter ranges from zero to one, inclusive,
and represents how coherent the population is,
with $\rho = 1$ for fully coherent and $\rho = 0$ fully disordered.

\subsection{Chimera-like index and metastability index}\label{sec:chimera-metastability-def}
To compare with a previous analysis~\citep{hizanidis2016chimera} of
chimera-like states \gls{celegans} models,
we define a pair of indices related to
chimera-like quality and metastability~\citep{shanahan2010metastable}.
First, we organize the game's nodes
into $M$ disjoint communities.
We split the nodes into $M=2$ communities $C_m$
by taking a subset
(the first \num{8E4} steps for computations ease)
of the simulation results
for the \gls{celegans} simulation with asymmetry $\alpha = \num{0.75}$.
We then calculate the covariance matrix $K_{i,j}$ of the strategy indices,
ensuring that communicative ($C$, $\phi_i$)
and non-communicative ($N$, $\phi_i$) are treated distinctly.
Next, we calculate the row-wise covariance-sums  $\sum_i K_{i,j}$
and form a community by collecting all nodes $j$ with covariance-sum
at least \num{1500}, \ie{} $\sum_i K_{i,j} \ge 1500$.
Note that \num{1500} is chosen as a high cutoff
to ensure we only group nodes with high covariance-sum;
for reference, \num{1500} is is approximately \SI{81}{\percent}
of the maximum covariance-sum.
Finally, we place all the remaining nodes in a second community.

With these disjoint communities $C_m$, we then calculate
the time-dependent, community-wise order parameter $\rho_m(t)$
as
\begin{equation}
  \rho_m(t) \coloneqq \frac{1}{N_m} \abs{\sum_{j \in C_m} e^{\im \phi_j}}
\end{equation}
across members $j$ of community $C_m$ with size $N_m$.
Then, we define a chimera-like index $\chi$
\begin{equation}
  \chi = \aab{
    \sigma_{\text{chi}}}_T
  \label{eq:chimera_index}
\end{equation}
where
\begin{equation*}
    \sigma_{\text{chi}} \coloneqq \frac{1}{M-1} \sum_{m=1}^M
    \pab{\rho_m(t) - \aab{\rho_m(t)}_M}^2
\end{equation*}
and a metastability index $\lambda$
\begin{equation}
  \lambda = \aab{
    \sigma_{\text{met}}}_M
  \label{eq:metastability_index}
\end{equation}
where
\begin{equation*}
    \sigma_{\text{met}} \coloneqq \frac{1}{T-1} \sum_{t=1}^T
    \pab{\rho_m(t) - \aab{\rho_m(t)}_T}^2
\end{equation*}
across the $M$ communities and $T$ time steps.
The chimera-like index measures the difference in coherence between communities:
complete homogeneity between communities
(\eg{} all fully synchronized \emph{or} fully disordered)
corresponds to $\chi = 0$,
while having $M$ communities
with half fully synchronized ($\rho_m = 1$)
and the other half fully disordered ($\rho_m = 0$)
for all times yields a maximum $\chi = M/[4(M-1)]
= 1/2$ for our $M=2$~\citep{shanahan2010metastable}.
Likewise, the metastability index $\lambda$ measures how metastable
the system is (\ie{} transiting between synchronicity and disorder).
A system that is fully synchronized or disordered gives $\lambda = 0$;
$\lambda$ is maximized for a system spending equal times in each state,
where the variance of the uniform distribution gives
$\lambda = 1/12 \approx \num{0.08}$~\citep{shanahan2010metastable}.

\backmatter{}

\bmhead{Supplementary information}
Supplementary Figure 1 of two-player game order graphs.
Section 1, two-player game order graphs;
Section 2, derivation of well-mixed communicative fraction
with symmetry breaking.
Supplementary videos 1-3,
time-evolution of \gls{celegans} player strategies
using the same color scheme as \cref{fig:graph}
with $B_0/c = 1.5$, $\beta_0/B_0 = 0.95$, $c = 0.1$,
$\mu = 0.0001$, $m = 20$, $\delta = 0.2$, and \num{8E6} time steps;
supplementary video 1 shows $\alpha = 0$,
supplementary video 2 shows $\alpha = 0.75$,
and
supplementary video 3 shows $\alpha = 1$.


\bmhead{Declarations}
The authors declare no competing interests.

\bmhead{Code availability}
An interactive webpage for exploring the data set is available at
\url{https://tzdyrski.github.io/egt-kuramoto/notebooks/EKT_Plots.html}.
Additionally, all source code is available via GitHub
at \url{https://github.com/TZdyrski/egt-kuramoto/tree/1.0.0}.

\bmhead{Data availability}
The processed data for all simulated parameter ranges
is available on Zenodo at \url{https://doi.org/10.5281/zenodo.17135745}.

\bibliography{references.bib}

\end{document}


\title{Supplementary Information for\\ 
``Evolutionary Kuramoto dynamics unravels origins of chimera states
in neural populations''}
\author{
Thomas Zdyrski, Scott Pauls, Feng Fu}

\maketitle

\tableofcontents
\section{Two-player game order graphs}\label{sec:order_graphs}
\begin{center}
    \includestandalone{tikz/game-payoffs}
    \captionof{figure}{}\label{fig:order_graphs}
\end{center}

\Cref{fig:order_graphs} shows the
order graphs for the 16 symmetric, ordinal two-player games.
Each sub-panel shows the payoff order for both the column
and row player.
The letters correspond to the following payoff matrix
\begin{equation*}
  \begin{bNiceMatrix}[first-col,first-row,hvlines]
    & \textcolor{blue}{C} & \textcolor{blue}{N} \\
    \textcolor{red}{C} & A & B \\
    \textcolor{red}{N} & C & D
  \end{bNiceMatrix}
\end{equation*}
The horizontal axis represents the order of row player payoffs,
and the red lines represent row player options.
The vertical axis represents the order of column player payoffs,
and the blue lines represent column player options.
The arrows show the direction of increasing
payoff for the relevant player,
and solid black dots represent Nash equilibria.
This figure is adapted from the topological taxonomy
order graphs~\citep{bruns2015names}.

\section{Well-mixed communicative fraction with symmetry breaking}\label{sec:analytic_comm_frac}
Here, we follow the steady-state communicative fraction derivation
in the previous EK study~\citep{tripp2022evolutionary}
and extend it to include a payoff asymmetry $\alpha$.
To keep the derivation tractable,
we only consider the well-mixed case.
We will assume a low mutation rate
so that any mutation fixates prior to the next mutation.
We then consider a population with one communicative species $E = (C, \phi_i)$
and one non-communicative species $F = (N, \phi_j)$.
Our first goal is to calculate
the (communicative) $\rho_E$ and (non-communicative) $\rho_F$
fixation probabilities of a single $E$ or $F$ invader, respectively.
First, we use our $\alpha$-dependent payoff matrix,
to calculate the $\alpha$-dependent average payoff
of each strategy across the entire population,
\begin{align}
  \pi_E(k) &= \ab(\frac{k-1}{n-1}) \pab{B_0 - c}
                + \ab(\frac{n-k}{n-1}) \ab(\beta(\Delta\phi) 2\alpha - c) \\
           &= \frac{1}{n-1}
               \ab(k \ab(B_0 - 2 \alpha \beta(\Delta\phi))
                 + 2 \alpha n \beta(\Delta\phi) - B_0 - (n-1) c)
\end{align}
and
\begin{align}
  \pi_F(k) &= \ab(\frac{k}{n-1}) \ab(\beta(\Delta\phi) 2 (1-\alpha))
                + \ab(\frac{n-k-1}{n-1}) \ab(0) \\
           &= \ab(\frac{k}{n-1}) 2 (1-\alpha) \beta(\Delta\phi)
\end{align}
Here, $k \in [0,n]$ is the number of $E$-players
$\Delta \phi$ is the phase difference between the $E$ and $F$
strategies,
and we defined $\beta(\Delta \phi) = \beta_0 f(\Delta \phi)$.
Note that the first term of $\pi_E$ corresponds
to the $E$-$E$ payoff, which has $\Delta \phi = 0$,
hence $B_0 f(\Delta \phi) = B_0$.
As mentioned in the main text,
each node has an exponential total fitness
$f_E = \exp(\delta (n-1) \pi_E)$
or
$f_F = \exp(\delta (n-1) \pi_F)$,
depending on its strategy,
with selection strength $\delta$.

In this paragraph, we summarize the relevant steps;
see the prior work~\citep{tripp2022evolutionary}
for more detailed derivations.
We now model the Moran process
as a Markov chain where each state $i \in \Bab{0,1,\ldots,n}$
is the number of (communicative) $E$ strategies.
Interestingly, we note that since the Moran process changes, at most,
one strategy per time step,
this Markov chain has the same fixation probabilities
as a (state-dependent) ``gambler's ruin'' problem with ties.
We denote $x_i$ as the probability that state $i$ will fixate
to state $n$ with all players using strategy $E$.
Then, the $E$ and $F$ fixation probabilities defined above are given by
$\rho_E = x_1$ and $\rho_F = 1 - x_{n-1}$.
Now, the Markov chain is described by the transition probability $p_{i,j}$
from state $i$ to state $j$ as
\begin{align*}
  p_{0,0} &= 1, \\
  p_{n,n} &= 1, \\
  p_{i,i-1} &= \frac{i}{n} \frac{(n-i) f_F(i)}{i f_E(i) + (n-i) f_F(i)} \\
  p_{i,i+1} &= \frac{n-i}{n} \frac{i f_E(i)}{i f_E(i) + (n-i) f_F(i)} \\
  p_{i,i} &= 1 - p_{i,i+1} - p_{i,i-1}
\end{align*}
Then, we obtain a recurrence relation for $x_i$ by conditioning
on the outcome of the first step:
\begin{align*}
  x_i &= x_{i-1} p_{i,i-1} + x_i p_{i,i} + x_{i+1} p_{i,i+1} \\
\end{align*}
with boundary values
\begin{align*}
  x_0 &= 0 \\
  x_1 &= 1
\end{align*}
Using the $p_{i,i} = 1 - p_{i,i+1} - p_{i,i-1}$ relation from above
to replace the $p_{i,i}$ term gives
\begin{equation*}
  \pab{x_i - x_{i-1}} \gamma_i = \pab{x_{i+1} - x_i}
\end{equation*}
Then, defining $y_i \coloneqq x_i - x_{i-1}$,
we find $y_1 = x_1$ and $y_{i+1} = \gamma_i y_i$,
yielding $y_i = \prod_{j=1}^{i-1} \gamma_j x_1$ for $i \ge 2$.
Finally, we form a telescoping sum to yield
\begin{equation*}
  1 - x_1 = x_n - x_1
  = \sum_{i=1}^{n-1} y_{i+1}
  = \sum_{i=1}^{n-1} \prod_{j=1}^i \gamma_j x_1
\end{equation*}
Solving this for $x_1$ gives
\begin{equation}
  \rho_E = x_1 = \frac{1}{1 + \sum_{i=1}^{n-1} \prod_{j=1}^i \gamma_j}
  \label{eq:comm_fixation_prob}
\end{equation}
Likewise, $x_i = \sum_{j=1}^i y_i = \sum_{j=0}^{i-1} y_{i+1}$, so
\begin{equation*}
  x_i
  = x_1 + \sum_{j=1}^{i-1} \prod_{k=1}^j \gamma_k x_1
  = \frac{1 + \sum_{j=1}^{i-1} \prod_{k=1}^j \gamma_k}
    {1 + \sum_{j=1}^{n-1} \prod_{k=1}^j \gamma_k}
\end{equation*}
Therefore, we find
\begin{equation*}
  \rho_F = 1 - x_{n-1}
  = 1 - \frac{1 + \sum_{j=1}^{n-2} \prod_{k=1}^j \gamma_k}
    {1 + \sum_{j=1}^{n-1} \prod_{k=1}^j \gamma_k}
  = \frac{\prod_{k=1}^{n-1} \gamma_k}{1 + \sum_{j=1}^{n-1} \prod_{k=1}^j \gamma_k}
\end{equation*}
This implies
\begin{equation}
  \frac{\rho_F}{\rho_E} = \prod_{k=1}^{n-1} \gamma_k
  \label{eq:fixation_prob_ratio}
\end{equation}
Therefore, we've found $\rho_E$ and $\rho_F$ for a population
with two fixed strategies.

Now, we need to derive the form of the $\gamma_k$,
which will include the new asymmetry factor $\alpha$:
\begin{align*}
  \gamma_k &= \frac{f_F(k)}{f_E(k)} \\
           &= \exp[\delta (n-1) (\pi_F(k) - \pi_E(k))]
           \\
           &= \exp\ab[\delta \ab(
    k 2 (1-\alpha) \beta(\Delta\phi)
    - k \ab(B_0 - 2 \alpha \beta(\Delta\phi))
                 - 2 \alpha n \beta(\Delta\phi) + B_0 + (n-1) c)] \\
      &= \exp\ab[\delta \ab(
    \ab(2 \beta(\Delta\phi) - B_0) k
                 + B_0 - 2 \alpha n \beta(\Delta\phi) + (n-1) c)]
\end{align*}
We also calculate
\begin{equation}
  \begin{aligned}
    \prod_{k=1}^j  \gamma_k
      &= \prod_{k=1}^j \exp\ab[\delta \ab(
      \ab(2 \beta(\Delta\phi_{qr}) - B_0) k
      + B_0 - 2 \alpha n \beta(\Delta\phi_{qr}) + (n-1) c)]
      \\
      &= \exp\ab[\delta \sum_{k=1}^j \ab(
      \ab(2 \beta(\Delta\phi_{qr}) - B_0) k
      + B_0 - 2 \alpha n \beta(\Delta\phi_{qr}) + (n-1) c)]
      \\
      &= \exp\ab[\delta \ab(
      \ab(2 \beta(\Delta\phi_{qr}) - B_0) \frac{j(j+1)}{2}
      + j \ab(B_0 - 2 \alpha n \beta(\Delta\phi_{qr}) + (n-1) c))]
      \\
      &= \exp\ab[\delta \ab(
      \ab(\beta(\Delta\phi_{qr}) - \frac{B_0}{2}) j^2
      + j \ab(\frac{B_0}{2} + \beta(\Delta\phi_{qr}) (1 - 2 \alpha n)  + (n-1) c))]
  \end{aligned}
  \label{eq:gamma_prod}
\end{equation}
Additionally, we note that substituting
\cref{eq:gamma_prod} into \cref{eq:fixation_prob_ratio}
with $j=n-1$ shows that $\rho_F/\rho_E$
depends of $\Delta \phi_{qr}$.
This is in contrast to the $\alpha=0.5$ case~\citep{tripp2022evolutionary}
where $\rho_F/\rho_E$ simplifies to the $\Delta \phi_{qr}$-independent form
$\rho_F/\rho_E = \exp \ab(\delta \ab[(n-1)c - B_0 (n-2)/2])$.

Next, we consider the long-time dynamics incorporating more than two strategies.
By again using the low-mutation-rate assumption,
we can assume that any mutation either fixates
or dies out before the next mutation.
Therefore, the system evolves from one homogeneous state to another
using the probabilities we just derived.
We can model this behavior with a new Markov chain
using the $m=20$ phases and two communicative strategies (C and N)
to give a $2m$-dimensional state space:
$\Bab{(C,\phi_1), \ldots (C,\phi_m), (N,\phi_1), \ldots (N,\phi_m)}$.
Then, the transition probability $\rho_{NC,\Delta \phi_{qr}}$
from state $(N,\phi_q)$ to state $(C,\phi_r)$ is just $\rho_E$ above.
Similarly, the probability $\rho_{CN,\Delta \phi_{rq}}$
that a $(C,\phi_r)$ state is successfully invaded by
an $(N,\phi_q)$ state is simply $\rho_F$ above.
By the rotation symmetry of $\phi$, each $\phi_i$ has the same probability of fixation.
Therefore, consider an arbitrary $\phi_q$ and denote its communicative fixation probability as
$s_1$ and its non-communicative fixation probability $s_2$.
We can calculate the ratio of these probabilities
by incorporating the fixation probabilities against all other $\phi_r$ phases:
\begin{align}
  \frac{s_2}{s_1} &= \frac
    {\sum_{r=1}^{m} \rho_{CN,\Delta \phi_{rq}}}
    {\sum_{r=1}^{m} \rho_{NC,\Delta \phi_{qr}} }
  \\
  &=
  \frac
  {\sum_{r=1}^{m} \rho_{NC,\Delta \phi_{qr}}
    \exp \Bab{
      \delta (n-1)
      \bab{
        (n-1) c + n \beta(\Delta \phi_{qr}) (1 - 2 \alpha)
        - \frac{n-2}{2} B_0
      }
    }
  }
  {\sum_{r=1}^{m} \rho_{NC,\Delta \phi_{qr}}}
  \\
  &=
  \bab{
    \frac
    {\sum_{r=1}^{m} \frac
      {\exp \Bab{ \delta (n-1)
        \bab{
          (n-1) c + n \beta(\Delta \phi_{qr}) (1 - 2 \alpha) - \frac{n-2}{2} B_0
        }}
      }
      {1 + \sum_{j=1}^{n-1} \exp\Bab{
        \delta \bab{
         \pab{\beta(\Delta \phi_{qr}) - B_0/2} j^2
         + j \pab{B_0/2 + \beta(\Delta \phi_{qr}) (1 - 2 \alpha n) + (n-1) c}
        }}
      }
    }
    {\sum_{r=1}^{m} \frac{1}
      {1 + \sum_{j=1}^{n-1} \exp\Bab{
        \delta \bab{
         \pab{\beta(\Delta \phi_{qr}) - B_0/2} j^2
         + j \pab{B_0/2 + \beta(\Delta \phi_{qr}) (1 - 2 \alpha n) + (n-1) c}
        }}
      }
    }
  }
  \label{eq:full_analytic_frac}
\end{align}
The first equality used \cref{eq:fixation_prob_ratio}
and the second equality used \cref{eq:comm_fixation_prob}.
Unlike in the $\alpha = 0.5$ symmetric case~\citep{tripp2022evolutionary},
the ratio $\rho_F/\rho_E$ depends on $\Delta \phi_{qr}$,
so we cannot factor the exponential component out of the sum
and cancel the $\rho_{CN,\Delta \phi_{qr}}$ terms.
However, we can asymptotically expand $s_2/s_1$ for small
$B_0/c \coloneqq \epsilon \ll 1$;
additionally, since all of our simulations use $\beta \propto B$,
we also assume $\beta/c \sim \epsilon \ll 1$.
Finally, to simplify the calculation,
we define the asymptotic expansion of
$\rho_{NC,\Delta \phi_{qr}} \coloneqq
\rho_{NC,\Delta \phi_{qr}}^{(0)}
+
\rho_{NC,\Delta \phi_{qr}}^{(1)}
+
\order{\epsilon^2}
$
where
$
\rho_{NC,\Delta \phi_{qr}}^{(i)} \sim \epsilon^i
$
depends only on terms of total order $i$ in $B_0/c$ and $\beta/c$.

Then, to first order in $\epsilon$
we have
\begin{align*}
  \frac{s_2}{s_1}
  &=
  \frac
  {\sum_{r=1}^{m} \rho_{NC,\Delta \phi_{qr}}
    \exp \Bab{
      \delta (n-1)
      \bab{
        (n-1) c + n \beta(\Delta \phi_{qr}) (1 - 2 \alpha)
        - \frac{n-2}{2} B_0
      }
    }
  }
  {\sum_{r=1}^{m} \rho_{NC,\Delta \phi_{qr}}}
  \\
  &=
  e^{\delta (n-1)^2 c}
  \Biggl\{
  \\
  &\qquad
  \frac
  {\sum_{r=1}^{m}
    \pab{
      \rho_{NC,\Delta \phi_{qr}}^{(0)}
      +
      \rho_{NC,\Delta \phi_{qr}}^{(1)}
    }
    \Bab{1 +
      \delta (n-1)
      \bab{
        n \beta(\Delta \phi_{qr}) (1 - 2 \alpha)
        - \frac{n-2}{2} B_0
      }
    }
  }
  {\sum_{r=1}^{m} \pab{
    \rho_{NC,\Delta \phi_{qr}}^{(0)}
    +
    \rho_{NC,\Delta \phi_{qr}}^{(1)}
  }}
  \Biggr\}
  \\
  &\qquad
  + \order{\epsilon^2}
  \\
  &=
  e^{\delta (n-1)^2 c}
  \Biggl\{
  \frac
  {\sum_{r=1}^{m}
    \pab{
      \rho_{NC,\Delta \phi_{qr}}^{(0)}
      +
      \rho_{NC,\Delta \phi_{qr}}^{(1)}
    }
  }
  {\sum_{r=1}^{m} \pab{
    \rho_{NC,\Delta \phi_{qr}}^{(0)}
    +
    \rho_{NC,\Delta \phi_{qr}}^{(1)}
  }}
  \\
  &\qquad
  +
  \frac
  {\sum_{r=1}^{m} \rho_{NC,\Delta \phi_{qr}}^{(0)}
    \delta (n-1)
    \bab{
      n \beta(\Delta \phi_{qr}) (1 - 2 \alpha)
      - \frac{n-2}{2} B_0
    }
  }
  {\sum_{r=1}^{m} \rho_{NC,\Delta \phi_{qr}}^{(0)}
  }
  \Biggr\}
  + \order{\epsilon^2}
  \\
  &=
  e^{\delta (n-1)^2 c}
  \Bab{
    1
    - \delta (n-1) \frac{n-2}{2} B_0
    +
    \frac{\delta n (n-1)}{m} (1 - 2 \alpha)
    \sum_{r=1}^{m} \beta(\Delta \phi_{qr})
  }
  + \order{\epsilon^2}
  \\
  &=
  \exp \Bab{\delta (n-1) \bab{
    (n-1) c
    - \frac{n-2}{2} B_0
    +
    \frac{n (1 - 2 \alpha)}{m}
    \sum_{r=1}^{m} \beta(\Delta \phi_{qr})
  }}
  + \order{\epsilon^2}
  \\
  &=
  \exp \Bab{ \delta (n-1) \bab{
    (n-1) c
    - \frac{n-2}{2} B_0
    +
    \frac{n (1 - 2 \alpha)}{2} \beta_0
  }}
  + \order{\epsilon^2}
\end{align*}
Where, in the last line, we used
the definition of
$\beta(\Delta \phi_{qr}) = \beta_0 \bab{1 + \cos(\Delta \phi_{qr})}/2$
to write
$\sum_{r=1}^m \beta(\Delta \phi_{qr})
= \beta_0 \sum_{r=1}^m \Bab{1 + \cos\bab{2 \pi (q-r)/m}}/2
= m \beta_0/2$
since the cosine sum gives zero.

Finally, since each $\Delta \phi_i$ is equally likely,
the probability of fixing to \emph{any}
communicative state is $m s_1$;
likewise the total non-communicative fixation probability is $m s_2$.
Then, since the process must eventually absorb to either $C$ or $N$,
we have
$m s_1 + m s_2 = 1$, giving the probability of communicative fixation as
\begin{equation}
  m s_1 = \frac{1}{1 + s_2/s_1}
  \label{eq:full_analytic}
\end{equation}
with $s_2/s_1$ given by \cref{eq:full_analytic_frac},
and the asymptotic, small-$\epsilon$ approximation given by
\begin{equation}
  m s_1 = \frac{1}{1 + s_2/s_1}
  \approx
  \frac{1}
  {1 + \exp \Bab{\delta (n-1) \bab{
    (n-1) c
    - \frac{n-2}{2} B_0
    +
    \frac{n (1 - 2 \alpha)}{2} \beta_0
    }}
  }
  \label{eq:analytic_first_term}
\end{equation}

\section{Data availability}
\subsection{Time-evolution animations}
The chimera-like quality of the games' populations
are necessarily time-dependent.
We found that animations showing the time-evolution
of players' strategies was a particularly useful way
to observe this chimera-like quality.
Therefore, we have included animations
showing the populations' evolving strategies
in the supplemental information for a subset of the parameters sampled.

\subsection{Interactive plots}
While discussing the simulation results in the main manuscript,
we necessarily focused on a particular region of parameter space.
Nevertheless, we also performed simulations across a range of parameters
to ensure the robustness of our results.
We used a Pluto.jl notebook to organize and visualize the data
and have made these results available for public analysis.
Additionally, this notebook includes time-evolution animations
for all of the simulated parameter combinations.

\bibliography{references.bib}